\documentclass[amsmath,amssymb,twocolumn,superscriptaddress,preprintnumbers,nofootinbib,prd]{revtex4-1}

\input{header}

\begin{document}

\preprint{IPPP/21/10, DESY-21-120}

\vspace*{-0.7cm}

\title{The potential of CMS as a high-energy neutrino scattering experiment}

 \author{P. Foldenauer}
 \email{patrick.foldenauer@durham.ac.uk}
 \affiliation{Institute for Particle Physics Phenomenology, Durham University, Durham DH1 3LE, United Kingdom} 
 \author{F. Kling}
 \email{felixk@slac.stanford.edu}
\affiliation{Theory Group, SLAC National Accelerator Laboratory, Menlo Park, CA 94025} 
\affiliation{Deutsches Elektronen-Synchrotron DESY, Notkestrasse 85, 22607 Hamburg, Germany}
\author{P. Reimitz}
 \email{peter@if.usp.br}
\affiliation{Instituto de F\'{i}sica,
Universidade de Sāo Paulo, 05508-090 Sāo Paulo, SP, Brasil}


\begin{abstract}
With its enormous number of produced neutrinos the LHC is a prime facility to study the behaviour of high-energy neutrinos. In this paper we propose a novel search strategy for identifying neutrino scattering via displaced appearing jets in the
high granularity calorimeter (HGCAL) of the CMS endcap in the high luminosity run of the LHC. We demonstrate in a cut-and-count based analysis how the enormous hadronic background can be reduced while keeping most of the neutrino signal. This paper serves as a proof-of-principle study to illustrate the feasibility of the first direct observation of high-energetic neutrinos coming from $W$ decays. 
\end{abstract}

\maketitle

\section{Introduction}\label{sec:intro}

The matter content of Standard Model (SM) consists of three generations of quarks and leptons. Of these the three neutral leptons, \ie the {\it neutrinos}, are arguably the most elusive. Being the lightest particles of the SM, not only are their masses and the mechanism responsible for generating them still undetermined, but they are also  very difficult to probe experimentally since they only participate in the weak interaction. Nevertheless, neutrinos have been detected from a vast variety of sources, including nuclear reactors, beam dumps, the atmosphere, the Earth, the Sun, supernovae and even extra-galactic astrophysical sources. 

In this study, we consider a novel complementary neutrino source, the Large Hadron Collider (LHC). As the highest-energy collider experiment, the LHC is also the source of the most energetic human-made neutrinos. However, the chance that those neutrinos scatter in any of the detector components is very low. To illustrate this, we can estimate the probability $P$ of a neutrino with energy $E_\nu$ to interact in a meter of material with density $\rho$ and obtain $P \sim 4 \cdot 10^{-13} \times [E_\nu/\gev] \times [\rho / (\text{g} \, \text{cm}^{-3} )]$. Therefore, neutrinos at the LHC have only been observed indirectly as missing transverse momentum. Nevertheless, as proposed already decades ago~\cite{DeRujula:1984pg, Vannucci:253670, DeRujula:1992sn, Park:2011gh, Feng:2017uoz, Buontempo:2018gta, Beni:2019gxv, XSEN:2019bel, Beni:2020yfy}, it is also possible in principle to directly observe LHC neutrinos through their scattering, despite the small interaction rate. Indeed, this idea was reconsidered in recent years. As a result, two new experiments with the goal to measure LHC neutrinos will start their operation during Run~3 of the LHC: FASER$\nu$~\cite{Abreu:2019yak, Abreu:2020ddv} and SND@LHC~\cite{Ahdida:2020evc, Ahdida:2750060}.  

\begin{figure*}[t]
\centering
\includegraphics[width=1\textwidth]{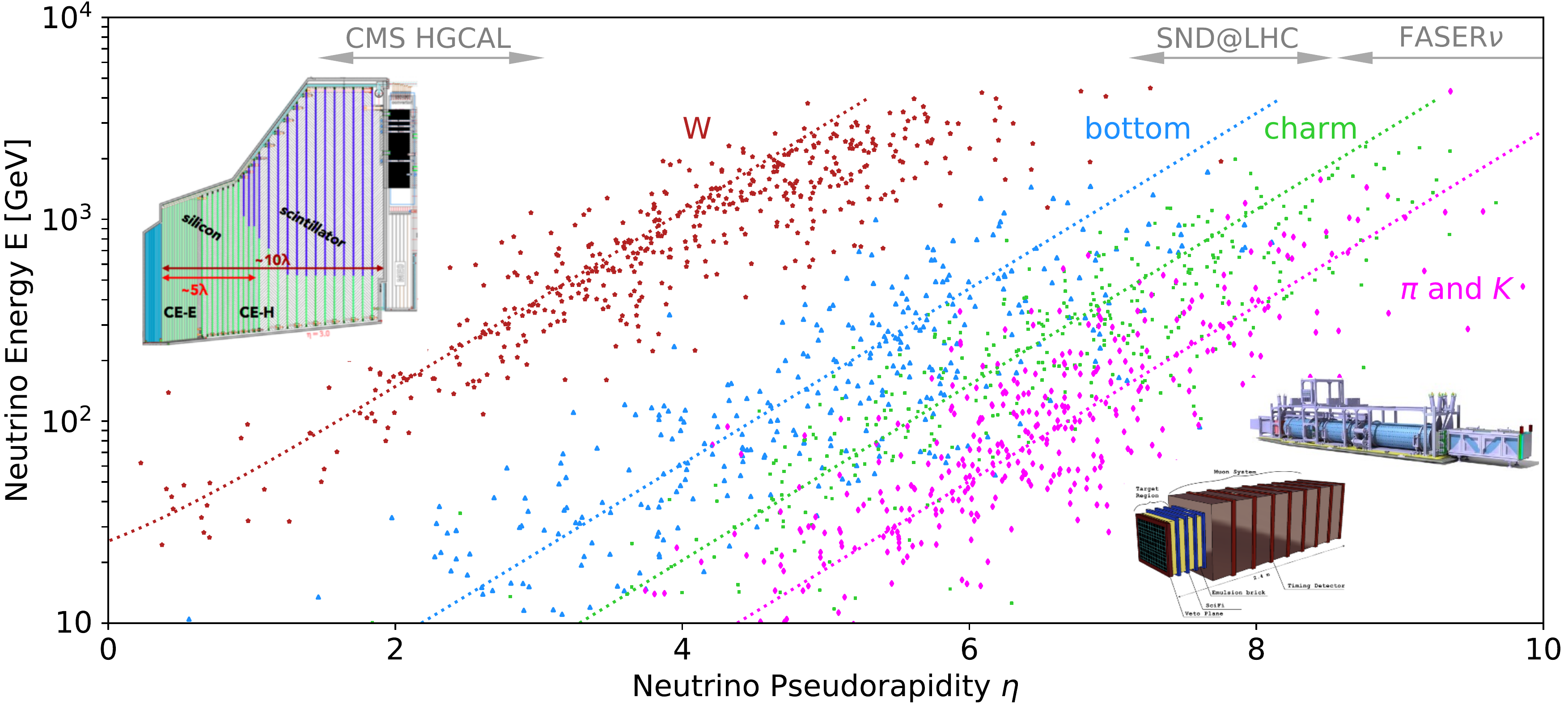}
\caption{\textbf{Neutrino Distribution}. The scattered points illustrate the distribution of neutrinos, as simulated using \texttt{Pythia~8} and weighted by the interaction probability, in terms of the neutrino pseudorapidity $\eta$ and the neutrino energy $E$. 
The differently colored populations denote neutrinos from the decay of $W$-bosons (red), bottom hadrons (blue), charmed hadrons (green) and light hadrons (magenta), where the number of points is the same for all production modes. The meta-stable pion and kaons are required to decay before colliding with the beam pipe or being deflected by the LHC magnets. The dashed lines correspond to the typical transverse momentum $p_T \sim m/2$ where $m$ is the mass of the parent particle. We also show the pseudorapidity coverage and illustrations of the CMS~HGCAL~\cite{collaboration:2017gbu}, and the dedicated forward LHC neutrino detectors SND@LHC~\cite{Ahdida:2020evc, Ahdida:2750060} and FASER$\nu$~\cite{Abreu:2019yak, Abreu:2020ddv}.
}
\label{fig:illustration}
\end{figure*}

Neutrinos at the LHC can originate from the weak decay of a variety of SM particles of different masses, ranging from pions to  electroweak gauge bosons. This is illustrated in \cref{fig:illustration}, where we show the distribution of neutrinos, weighted by their interaction probability $P$, in terms of their rapidity and energy. We can see that the population of neutrinos cluster around $p_T \sim m/2$, where $m$ is the mass of their parent particle, as denoted by the dotted lines. Therefore, depending on the detector location and probed energy range, different production modes can dominate the neutrino flux. 

Both FASER$\nu$ and SND@LHC are emulsion based detectors located in the far-forward direction, corresponding to large rapidities, about $480~\text{m}$ downstream from the ATLAS interaction point. At this location, they can be placed on or close to the beam collision axis, where the rate of neutrino interactions per detector volume is maximized. Both experiments use an emulsion based technology. In the case of FASER$\nu$, the emulsion detector is followed by a magnetized spectrometer, while for SND@LHC the emulsion is interleaved with tracking layers and followed by a muon system. The feasibility of neutrino detection using emulsion detectors at this location was recently demonstrated by the FASER collaboration, which reported the observation of the first neutrino interaction candidates at the LHC~\cite{Abreu:2021hol}. Both experiments will mainly detect neutrinos from pion, kaon and charm decays, and their corresponding neutrino fluxes have recently been discussed in Ref.~\cite{Kling:2021gos}. 

In \cref{fig:illustration}, we also include an ensemble of neutrinos from $W$-boson decay which populate a more central rapidity range. These neutrinos can scatter in the calorimeters of the LHC's large multipurpose detectors, ATLAS or CMS, and leave observable energy deposits. The main challenge is to distinguish those neutrino scattering events from the scattering of other long-lived neutral particles, such as neutrons or neutral kaons, whose production and interaction rates are many orders of magnitude larger. 

In this study, we present an analysis strategy to  separate the neutrino signal from hadronic backgrounds, which demonstrates that a search for neutrino interactions in the main LHC detectors can be performed in principle. We will focus on a particularly promising neutrino target, the high granularity calorimeter (HGCAL) upgrade of the CMS endcap~\cite{collaboration:2017gbu}. As we will see, it provides both a fine spatial segmentation and pile-up-rejection capabilities, which will play a crucial role in suppressing the hadronic backgrounds.

The rest of the paper is organized as follows: In~\cref{sec:technical} we give a brief description of the future CMS HGCAL and its unique spatial and temporal resolution. In~\cref{sec:nuscattering} we lay out the search strategy and discuss the relevant analysis-level cuts necessary for detecting neutrino scattering in the HGCAL. Finally, in~\cref{sec:results} we summarise our results and give a brief outlook on the future potential of this type of search strategy.

\section{The CMS high granularity calorimeter}\label{sec:technical}

In this section, we want to  briefly summarise the main characteristics of the future CMS HGCAL.

\textbf{Detector Geometry and Design -} During the high-luminosity run of LHC (HL-LHC) one of the main challenges for LHC detectors are backgrounds from the very high number of collisions occurring in the same bunch crossing - the {\it pile-up}. In order to discriminate interesting events from background, LHC detectors will therefore need excellent spatial and temporal resolution. This requirement, together with the fact that the expected fluence of neutrons and the level of ionizing radiation is significantly higher than in current runs (in particular close to the beam pipe) has led  to  a novel design of the CMS endcap calorimeter for the HL-LHC~\cite{collaboration:2017gbu, Rusack:2239181}.

The planned upgrade of the CMS endcap calorimeter - the HGCAL -  is a sampling detector located at $|z|=3.2$ m from the interaction point (IP) and extending to $|z|=5.2$ m. It covers the pseudo-rapidity range $1.5<|\eta|<3.0$ and  mostly utilizes silicon as the active material. One of the main arguments for silicon is its good performance under the high expected radiation levels with neutron fluences of up to $10^{16}$ n$_\mathrm{eq}/\mathrm{cm}^2$  and doses of 2 MGy expected for the HL-LHC. The HGCAL consists in total of three different segments~\cite{collaboration:2017gbu}. The first segment closest to the IP is the electromagnetic calorimeter (ECAL) consisting of a silicon-tungsten detector with 28 sampling layers totaling $26\, X_0$ radiation lengths and $1.7\, \lambda$ hadronic interaction lengths. The individual silicon cells have a size of 0.5 cm$^2$ - 1 cm$^2$ in the transverse plane. This finely granulated detector design allows a high-resolution measurement of the lateral development of electromagnetic showers. This helps to achieve a good two-shower separation  and the observation of narrow jets. For example, for photon showers of $p_T>40$ GeV a resolution of better than 4 mrad will be achievable~\cite{collaboration:2017gbu}.
Following the electromagnetic calorimeter comes a 24-layer hadronic sampling calorimeter (HCAL) of $\sim 8.5\, \lambda$ length, which employs two different detector designs. The first HCAL  segment is a  silicon-stainless steel sampling detector 
that has the same transverse cell size as the ECAL.  The second segment of the HCAL consists of a plastic scintillator-stainless steel sampling detector
with a transverse cell size of 4 cm$^2$–  30 cm$^2$. In total the HGCAL has a hadronic interaction length of $\sim 10\, \lambda$, such that most of the hadronic shower is contained in the detector and due to the finely granulated segmentation can be observed with high resolution, both in the transversal and longitudinal direction.

Since we are interested in the measurement of neutrino scattering, the main issue will be the mitigation of neutral hadronic backgrounds. Due to its excellent spatial and temporal resolution, the CMS HGCAL is the best candidate for this task at the LHC. In particular the excellent timing capabilities will be crucial in reducing the amount of neutral background from pile-up as we will discuss in a moment. For comparison, the ATLAS liquid argon barrel and endcap calorimeters~\cite{CERN-LHCC-2017-018} have a coarser spatial segmentation leading to a worse angular resolution. Furthermore, these calorimeters themselves do not provide any timing information, which makes suppression of neutral hadronic backgrounds from pile-up extremely challenging. The CMS barrel calorimeter does provide timing information, but only in the ECAL part of the detector~\cite{CERN-LHCC-2017-011}. Furthermore, the spatial granularity of the detector is inferior to the HGCAL. Both ATLAS~\cite{CERN-LHCC-2020-007} and CMS~\cite{CMS:2667167} are planning to install separate timing detectors in their Upgrade-II before the HL-LHC run. However, this will only help to mitigate pile-up from charged tracks and will not help reduce backgrounds from neutral hadrons. Similarly, the LHCb collaboration has plans for incorporating a dedicated timing detector in front of the ECAL~\cite{Aaij:2244311}, which again only allows for timing of tracks. Furthermore, LHCb will collect significantly less luminosity than ATLAS or CMS, making the observation of neutrinos unfeasible due to the small interaction rates.

\textbf{Pile-up Mitigation with Timing -} When analysing neutrino scattering events in the calorimeter during the high-luminosity run of the LHC, care has to be taken to eliminate potential pile-up contamination.

In this context, the CMS endcap HGCAL will be of special importance, since it will be able to mitigate pile-up significantly due to its timing capabilities~\cite{collaboration:2017gbu}. E.g. neutral~$K_L^0$ with $p_T> 5$ GeV are reconstructed with an efficiency of $>90\%$  and a resolution of $\lesssim 30$ ps.

The typical timing window chosen to remove pile-up is 90 ps~\cite{collaboration:2017gbu} and corresponds to a path length difference traveled by relativistic particles of $\Delta_l \sim 2.7$ cm. In order to further enhance the effect of this timing window on pile-up mitigation, a novel collision technique for the HL-LHC has been proposed, the so-called {\it crab kissing}~\cite{Fartoukh:2014nga}. In the crab-kissing scheme, the bunches are colliding at a crossing angle. However, before the collision they pass an RF cavity giving the tail and head of the bunches a kick so that they rotate and collide with maximum spatial overlap~\cite{Verdu-Andres:2016psv}. This way the distance over which the various pile-up collisions occur is stretched out in space and can be described by a supergaussian distribution of order 4~\cite{MedinaMedrano:2301928}. The pile-up density at the HL-LHC in the optimized case (crab kissing) has a typical spatial extension (full width at half maximum) of 31.4 cm with the nominal number of expected pile-up events being about $130$ per bunch crossing~\cite{MedinaMedrano:2301928}. 

If we approximate the supergaussian distribution in the crab kissing scenario by a uniform distribution, the timing resolution will allow us to reduce the number of pile-up events per bunch crossing on average to roughly\footnote{For simplicity we have assumed that the path length difference of $\Delta_l\sim 2.7$ cm corresponds to the difference of the vertex position in the z-direction. However, since the HGCAL is not positioned exactly along the z-axis, the vertex separation is obtained by $\cos \theta \, \Delta_l$, where $\theta$ is the angle of emission of the particle. Nevertheless, e.g.~in the center of the angular coverage at $\eta\sim 2.3$ we have $\cos\theta \sim 0.98$, which can safely be ignored.}
\be
    N_{\rm pu} \approx \frac{2.7}{31.4} \ 130 \sim 11 \,.
\ee
While in reality, the observed number of pile-up events follows a distribution, in this study we will for simplicity assume a constant number of pile-up events of  $N_{\rm pu}=10$.

\section{Neutrino scattering at LHC}\label{sec:nuscattering}

Highly energetic neutrinos are abundantly produced in the primary $pp$ collisions at the LHC. However, their probability to interact with the main detectors is very low, and they their presence only manifests itself as missing transverse momentum. This could change at the HL-LHC, where their abundance is sufficiently large such that a sizable number of neutrinos are expected to scatter with nuclei in the denser part of the detector, \eg the calorimeter. To be more precise, we expect $\mathcal{O}(10)$ muon neutrinos from $W$-boson decays to interact with the hadronic calorimeters of ATLAS and CMS. In this work we explore the potential of observing such neutrino scattering events with LHC multipurpose detectors in a cut-and-count based search strategy. 

\textbf{Signal and Background Characteristics -}
In the following analysis a signal event is defined as a neutrino scattering with the detector material. The scatterings occur mostly via the charged current interaction at high energies, where the neutrino converts into a charged lepton and the scattered quark appears as a jet. This provides for a characteristic signature to be searched for in neutrino scattering at the LHC, where a charged lepton and a single energetic jet are produced in a single displaced vertex in the calorimeter with no tracks pointing back to the IP.

As seen in \cref{fig:illustration}, $W$-boson decays are a promising source of highly energetic neutrinos with $E \gtrsim 10^2$~GeV for interactions within the CMS HGCAL. In the SM we expect approximately equal decay rates into $W \!\to\! e\,\nu_e$, $\mu\,\nu_\mu$ and $\tau\,\nu_\tau$ final states at high energies. However, due to the excellent muon reconstruction efficiency of CMS we focus on the muonic decay of the $W$-boson as the most promising signature. The signal is hence given by the subsequent process,
\be
 \text{Production:} &\quad 
 qq' \to W\to \mu_1~\nu_\mu\,,  \\
\text{Scattering:} &\quad
\nu_\mu N \to \mu_2 + \text{jet}  \,,
\label{eq:signal}
\ee
{with a primary muon $\mu_1$ coming from the production process and a secondary muon $\mu_2$ in the scattering.}
To determine the neutrino interaction rate inside the detector material, we use the DIS neutrino interaction cross-sections evaluated at leading order as presented in Ref.~\cite{Abreu:2019yak} (for a recent reviews on neutrino cross sections see Refs.~\cite{Formaggio:2012cpf, McFarland:2008xd}). Assuming a dominant quark over anti-quark contribution in the initial state, the neutrino cross section scales as $d\sigma/(dx\,dy)\sim 1$ and the anti-neutrino cross-section scales as $d\sigma/(dx\,dy) \sim (E_{\mu_2}/E_\nu)^2$. This leads to a larger overall scattering rate of neutrinos compared to anti-neutrinos.

\begin{figure*}[t]
\centering
\includegraphics[width=.49\textwidth]{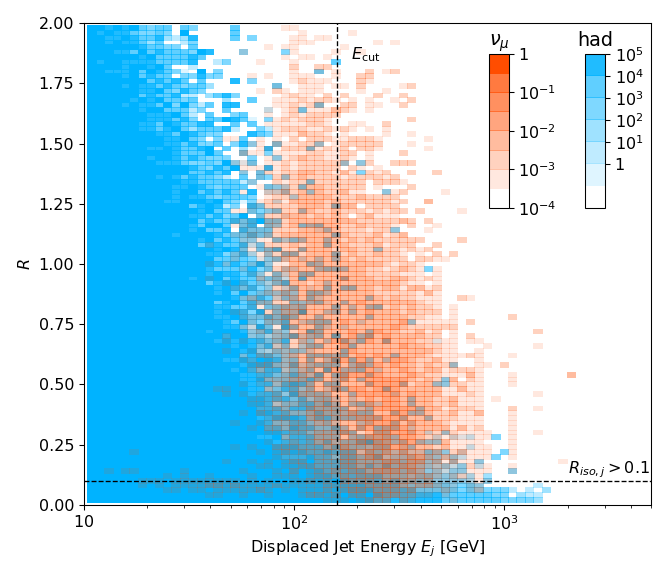}
\includegraphics[width=.49\textwidth]{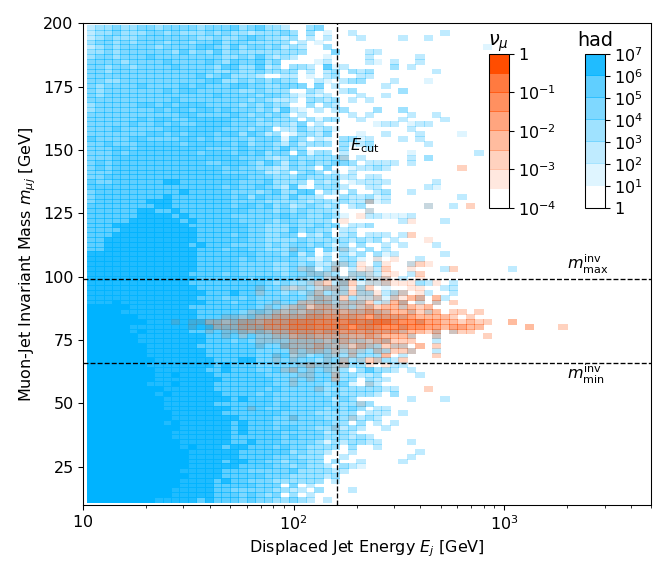}
\caption{\textbf{Left:} Distribution of signal (red) and background (blue) events in the jet energy versus $R$ plane after applying the isolated muon cut. All events below dashed horizontal line and left of the dashed vertical line will be rejected in the analysis by the jet isolation $R_{{\rm iso},j}>0.1$ and energy cut $E_{\rm cut}>160$~GeV, respectively. \textbf{Right:} Signal (red) and background (blue) histograms in the jet energy versus muon-jet invariant mass plane after applying the isolated jet cut. The two dashed horizontal lines indicate the window of selected events after the W mass cut $66~\gev < m_{\mu\nu}<99~\gev$.  
}
\label{fig:deltaR}
\end{figure*}

The dominant backgrounds to this process are  isolated muons accompanied by a long-lived neutral hadron interacting with the detector material. Possible candidates to fake the displaced jet of a neutrino signal are neutrons, $K_L^0$'s, $\Lambda$'s, or $\Xi^0$. Those neutral hadrons are abundantly produced in hard scattering, the underlying event or can come from pile-up and are very likely to interact with the detector material. The leading backgrounds including an isolated muon along with these hadrons are
\begin{itemize}
    \item[i)] The production of a leptonically decaying $W$-boson in association with additional neutral hadrons:
    \be
       qq' \to W + {\rm QCD}  \to \mu_1 + {\rm QCD} \,,
    \ee
    \item[ii)] Heavy quark production in association with additional neutral hadrons, where a muon is produced in heavy hadron decay:
    \be
       qq' &\to b/c + {\rm QCD}  \to  \mu_1 + {\rm QCD}\,.
    \ee
\end{itemize}
In both cases, the displaced jet is caused by the subsequent scattering of a neutral hadron in the back parts of the calorimeter:
\be
    \text{neutral hadron} + N &\to \text{jet}.
    \label{eq:bkg0}
\ee

In this study, we neglect further subleading sources of muons, such as the decays of $Z$-bosons or in-flight decays of light long-lived hadrons.
{In the following parton-level based analysis the label $j$ is used for the final-state jet-muon system in the scattering process. Likewise we use the term jet for the total final-state particle content. As an example, $E_j=E_{\mu_2}+E_{\rm had}$ denotes the energy of the incident particle or, equivalently, the total displaced jet energy and $E_{\rm had}$ is the energy carried by hadronic particles produced  in  the scattering process. In reality, we expect that a typical scattering produces a cone of $\mathcal{O}(10)$ particles that defines the jet~\cite{Ismail:2020yqc}. In a more refined analysis including a full detector simulation, we would, then, have to specify a jet-finding algorithm in order to define the jet.}

\begin{table}[t]
\centering\renewcommand\arraystretch{1.3}
\begin{tabularx}{.48\textwidth}{X | X | X }
\hline
\hline
 Cuts & Hadrons & Neutrinos\\
\hline
isolated muon & $1.02\cdot 10^{11}$ & 7.59 \\
isolated jet & $8.63\cdot 10^{10}$ & 7.05 \\
$W$ mass  &  $1.92\cdot 10^9$ & 6.55 \\
secondary muon  & $3.49 \cdot 10^5$  & 5.48 \\ \hline
$E_\mathrm{j}> 160$ GeV &  3.52 & 3.60 \\
\hline
\hline
\end{tabularx}
\caption{Cut flow table with number of neutral hadron background and neutrino signal events for the cut selection $R_{{\rm iso},\mu_1} \!>\! 0.1$, $R_{{\rm iso},j} \!>\! 0.1$, $m_\mathrm{inv} \!\in\! [66,99]$, $E_{\mu_{2}}/E_{\rm j} \!>\! 0.33$ and $E_\mathrm{j} \!>\! 160~\gev$. See text for details.
}
\label{tab:cutflow}
\end{table}

\textbf{Simulation -} Since the goal of this study is to demonstrate the feasibility of detecting neutrino scatterings at the LHC, we perform all simulations for a data set of 3 ab$^{-1}$ at leading order including parton shower and hadronization. In particular, all signal and background event samples are simulated with \textsc{Pythia~8.2}~\cite{Sjostrand:2007gs, Sjostrand:2014zea}. As discussed in Sec.~\ref{sec:technical}, we include a constant number of $N_{\rm pu}=10$ pile-up events for every Monte Carlo event. In this study, no full detector simulation was applied. However, we note that the cuts applied in this analysis are chosen conservatively to account for the anticipated detector resolutions. For example, the $R$ separation cuts as used in this paper are chosen large enough to be resolved by the spatial 2D segmentation of the high-resolution CMS HGCAL. 

\textbf{Analysis Strategy -} The events of interest consist of one muon and one displaced jet appearing the hadronic calorimeter. All generated samples for signal and background require a muon in the range $|\eta|<4$, according to the planned updated tracker and muon detector at CMS~\cite{Cepeda:2019klc, CERN-LHCC-2017-009, CERN-LHCC-2017-012}. To reduce background from hadronic activity, we discard any displaced jets occurring in the first radiation length of the HCAL. Therefore, we require the displaced jet to appear in the endcap HCAL ($1.5<|\eta|<3$) at 2 to 10 interaction lengths. 

The further analysis strategy is presented in \cref{tab:cutflow}. All cuts in our cut-and-count analysis are set in a very basic way and leave room for improvements in future searches for high-energy neutrinos. We start with a trigger muon requirement. It is followed by applying a jet isolation cut. Since all signal events are coming from $W$ production, we require the neutrino and muon to reconstruct the $W$ mass. Furthermore, we use the fact that a neutrino scattering process is producing a secondary muon. Finally, we focus on the high-energy tail of the spectrum to enhance the signal-to-background ratio.

\bigskip

\paragraph{Isolated central muon:} As a first basic cut, we require an isolated central muon with
\be
R_{\text{iso},\mu_1}>0.1, \quad p_{T,\mu_1}>20~\text{GeV}, \quad |\eta_{\mu_1}|<2.4~.
\ee
{Throughout the analysis, we only consider hadrons and charged leptons with $E > 10$~GeV in the calculation of the separation $R=\sqrt{\Delta\eta^2+\Delta\phi^2}$ of two particles where $\phi$ is the azimuthal angle about the LHC beam axis. If a certain particle is required to be isolated, its distance $R_{\rm iso}$ to any hadron or charged lepton has to be greater than a certain specified value.}  In the hard QCD production of $c\bar{c}$ and $b\bar{b}$ quark pairs, the muons are produced in the decay of the heavy hadron, mostly forward with low $p_T$ and accompanied by another meson from the same decay of B and D mesons. Hence, the muon trigger requirement, $p_{T,\mu_1} \!>\! 20$~GeV and $|\eta_{\mu_1}|<2.4$, as well as the isolation are reducing the amount of background events significantly.

\bigskip

\paragraph{Isolated jet:} The signal scattering event can be faked by neutral hadron scatterings in the detector. Here we use that energetic neutral hadrons are mostly produced as part of an energetic, and hence collimated, hadronic shower. To distinguish between neutral hadrons and neutrinos, we therefore apply a jet isolation cut,
\be
    R_{{\rm iso},j}>0.1\, .
\ee
As can be seen in \cref{tab:cutflow}, this does not have a large effect on the overall number of background events, which is dominated by low energy events. However, the number of background events in the relevant high energy bins is reduced significantly as seen in the left panel of \cref{fig:deltaR}. This figure also indicates that an energy-dependent $R$ cut could be applied as well to improve the signal/background separation even further.  

\bigskip

\begin{figure*}[t]
\centering
\includegraphics[width=.49\textwidth]{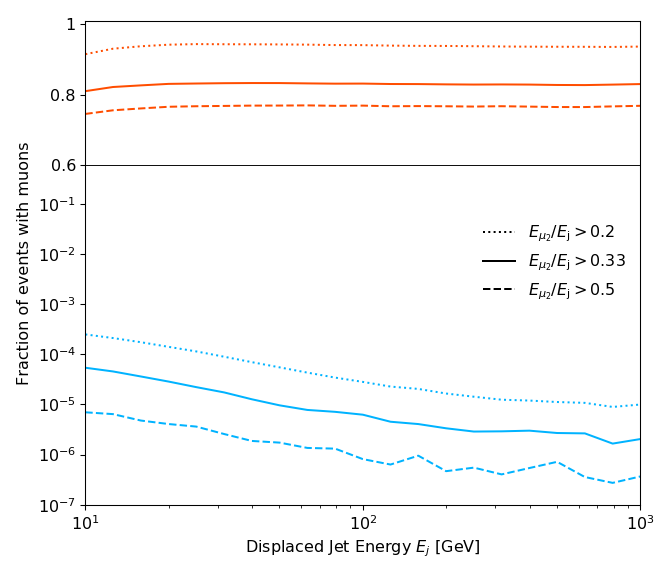}\hfill
\includegraphics[width=.49\textwidth]{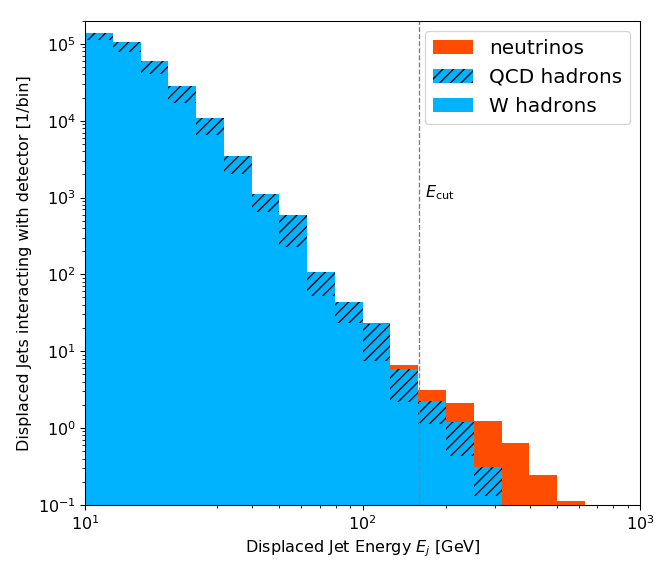}
\caption{\textbf{Left:} Fraction of muon neutrino interaction events (red) and neutral hadron interaction events (blue) containing a muon which carries more than 20\% (dashed), 33\% (solid) and 50\% (dashed) of the energy, as function of the displaced jet energy. \textbf{Right:} Neutrino scattering events (red) and neutral hadron events (blue) after all cuts. The vertical line illustrates the energy cut applied in our analysis.}
\label{fig:sigbkg}
\end{figure*}

\paragraph{W mass cut:} Additionally, we require the muon and neutrino jet to reconstruct the $W$ mass. Therefore, we set a cut on their invariant mass inspired by the ATLAS $W$-boson search~\cite{Aaboud:2017svj},
\be
66~\gev < m_{\mu\nu} < 99~\gev \,,
\ee
in the muon-neutrino system where $(p_{\mu_1}+p_\nu)^2=m_{\mu\nu}^2$. As seen in the right plot of \cref{fig:deltaR}, this encloses the neutrino signal events nicely whereas a large part of the background is removed.

\bigskip

\paragraph{Secondary muon:} Besides different incident scattering particles, another distinguishing feature between the processes in \cref{eq:signal} and \cref{eq:bkg0} is the {presence of a} secondary muon produced in the neutrino scattering process. We can use the additional muon as an extra handle to isolate the signal, and search for it with the CMS endcap muon stations located behind the HGCAL covering $1.2 \!<\! |\eta| \!<\! 2.4$~\cite{Sirunyan:2018fpa}. Unlike the trigger muon, the secondary muon is a standalone muon track not matching with any tracker track. 
We refrain from including efficiency factors and leave this to a detailed detector simulation. Besides, as will be explained below, we will apply an energy condition for the secondary muon which reduces the possibility of mis-identification by low-energy neutral hadrons even further. 

{Charged current muon neutrino interactions almost always produce an energetic secondary muon. However, in some rare cases energetic secondary muons can also be produced in the interactions of neutral hadrons with the calorimeter, for example via the in-flight decay of pions produced in these collisions, causing a background for the neutrino interaction signal. To determine the fraction of events, both for the muon neutrino interaction signal and the neutral hadron interaction background, that contains an energetic secondary muons, we have performed a separate simulation using \textsc{Pythia~8}~\cite{Sjostrand:2007gs}. We have done this by simulating the collisions of these particles with a fixed iron target, using the \textsc{nCTEQ~15}~\cite{Kovarik:2015cma} parton distribution function for iron, at different incident particle energies between $10~\gev$ and $1~\tev$. Neutral hadrons are represented by neutrons, which contribute the majority of neutral hadrons in this analysis, and about 30 million events are generated for each beam energy.}
{For simplicity, we have assumed that these secondary muons are detected with 100\% efficiency in the CMS muon endcap muon station. In a more detailed study one would have to perform a fully-fletched detector simulation for the charged current neutrino interactions. Such a study is also necessary to investigate any additional possible backgrounds from fake muons. For the purpose of this proof-of-principle study, however, we assume that this background is negligible.} 

For neutral hadron collisions with energies below $E \!<\! 100~\gev$, muons are mainly produced in decays of pions and kaons, which we require to decay within one nuclear interaction length $\lambda_{\text{int}} = 18.5~\cm$. Above $E \!>\! 100~\gev$, the secondary muons in background events are dominantly produced in prompt decays, for example from $\eta$, $\omega$, and $\phi$ mesons. Charm decays seem to constitute a sub-dominant component. Contrary to muons of neutrino interactions, most muons produced in the interactions of neutral hadrons with the HGCAL are soft and carry away only a small fraction of the incident particle energy. 

The left panel of~\cref{fig:sigbkg} shows the fraction of events where muons obtain 20\%, 33\%, or 50\% of the energy of the initial particle. We can see that the fraction of signal events with secondary energetic muons remains well above 70\% for all cut scenarios. We note that anti-neutrino interactions lead to a larger muon energy fraction compared to neutrino interaction, due to the additional $(E_{\mu_2}/E_\nu)^2$ factor in the scattering cross section mentioned earlier. In comparison, the amount of events in hadronic processes is suppressed significantly if the produced muon is required to inherit a substantial fraction of the initial energy. Hence, we require the hardest muon originating from the detector material collision to carry away 
\be
    E_{\mu_2}/E_\mathrm{j}>0.33\,.
    \label{eq:secondary}
\ee
With a fraction of secondary muons originating from neutral hadron-detector material collisions of roughly $10^{-4}$, this turns out to be a very effective handle to suppress neutral hadron backgrounds. 

\bigskip

\paragraph{Energy cut:} Neutrinos coming from $W$-bosons are expected to be more energetic while neutral hadron background tend to be softer, especially after requiring the displaced jet to be isolated. Hence, we only consider events with displaced jet energies above $E_{\rm cut}>160~\gev$ in order to increase the number of signal events over background events.

\bigskip

Starting with about $10^{11}$ background events and $\mathcal{O}(10)$ neutrino events at the muon trigger-level cut, we finally obtain comparable rates after all cuts. While reducing the number of events only mildly in the signal region, we can almost entirely erase the background in the signal region above $E_{\rm cut} > 160$~GeV.  
In the right plot of \cref{fig:sigbkg} we present the results after all cuts except for the energy cut. Neutral hadron interactions coming from $c\bar{c}$ and $b\bar{b}$ events are labeled as QCD hadrons. The background smoothly decreases towards higher displaced jet energies. This makes it easy to subtract the background from the signal events accumulating at the high energy tail of the distribution. 
To sum up, the basic cut-and-count analysis presented in this paper shows that there is the potential to see high-energy neutrino scatterings at the LHC.
Despite only having a very limited amount of neutrino scattering events inside the CMS HGCAL, the large hadronic background can be very successfully reduced in a way that signal events dominate the high-energy tail of the displaced jet energy spectrum. Already with very basic cuts as set in this paper, we count approximately an equal number of $\sim 4$ signal and background events above $E_j > 160~\gev$.

\textbf{Sources of Systematic Uncertainties -} {Since the analysis performed in this proof-of-principle study mostly relies on truth-level MonteCarlo events generated with \texttt{Pythia}, there are a number of potential sources of systematic uncertainties we have to take into account. Focusing on the theoretical description and neglecting any purely experimental effects, we consider the following list of systematic uncertainties:}

\textit{Primary Interaction:}
{A first class of uncertainties is associated with the simulation of the primary interaction, so the production of $W$-bosons as well as charm and bottom quarks. Here uncertainties arise for example from the choice of scales or the parton distribution functions, and can affect the production rates and kinematic distributions. However, the $W$-boson~\cite{CMS:2020cph}, charm~\cite{LHCb:2015swx, CMS:2021lab} and bottom~\cite{CMS:2016plw, LHCb:2017vec} production have been constrained by measurements already, which can be used to estimate and reduce the uncertainties in a data-driven way.}

\textit{Hadronization:} {An additional source of uncertainties is associated with the modelling of hadronization. This will have an important impact on the hadron distributions and can affect the effectiveness of the isolation cuts, especially for the isolated jet requirement for neutral hadron backgrounds. One approach would be to use tuning uncertainties~\cite{Buckley:2018wdv}, as implemented for example in the \textsc{ATLAS~A14} tune~\cite{ATL-PHYS-PUB-2014-021}, to estimate the modelling uncertainties. Alternatively, one could use a data driven approach and for example directly validate the $R_{\text{iso},\mu_1}$ distribution for muons from $W$ decay which should be the same as the $R_{\text{iso},j}$ distribution for neutrinos.}

\textit{Secondary Interaction:}
{There are further uncertainties associated with the modelling of the secondary interactions, for example on the production rates and kinematics of the produced hadrons as well as muons produced in subsequent decays. In particular, this could affect the effectiveness of the secondary muon cut in \cref{eq:secondary}, which was found to be a powerful handle to suppress the background. While we have used \textsc{Pythia~8} to determine the fraction of events with energetic muons, one could use other dedicated hadronic generators such as \textsc{EPOS}~\cite{Pierog:2013ria}, \textsc{SIBYLL}~\cite{Riehn:2019jet}, \textsc{QGSJET}~\cite{Ostapchenko:2010vb} or \textsc{DPMJET}~\cite{Roesler:2000he, Fedynitch:2015kcn} to estimate the uncertainties associated with the modelling of these collisions. These generators have been tuned to relevant fixed target data, for example from NA61~\cite{NA61SHINE:2017fne}, and would ideally be included in a full \textsc{Geant~4}~\cite{Agostinelli:2002hh} simulation to accurately model muons from in-flight pion and kaon decays. It might also be possible to constrain the muon rate in a data-driven way, either using fixed target experiments as suggested in Ref.~\cite{VanHerwijnen:2213605} or directly via an analysis of hadronic shower data in the CMS calorimeters.}

{Finally, a fully data-driven way to verify our background estimates would be to perform a sideband analysis. For example, as can be seen in the right panel of~\cref{fig:deltaR}, the muon-jet invariant mass $m_{\mu j}$ distribution outside the signal window  $66~\gev < m_{\mu j}<99~\gev$ would be ideally suited for estimating the backgrounds and interpolating them to the signal region.}

\textbf{Alternative Search Strategy -} An orthogonal way to search for neutrino scattering events is in heavy flavour production (i.e.~$c\bar{c}$ and $b\bar{b}$ events), where the neutrinos are produced in the decay of heavy meson.
For example, two such promising candidate are the decays $D^0 \to K^- \mu^+\, \nu_\mu$ and $B^0\to D^-\mu^+\, \nu_\mu$. Such decays can typically be identified using heavy flavour tagging techniques which use the presence of a reconstructed secondary vertex due to the comparably long lifetime of $B$ and $D$ hadrons. However, one  major difference to the search for neutrinos from $W$ production is that the produced muon is not isolated anymore. On the contrary, one expects to find other particles originating from the heavy hadron decay close by. For example, in the case of $D^0$ decay mentioned above, the typical angular separation scales as $R \sim m_{D} / p_{T,D} \sim 2 / 40 \sim 0.05$. This is a major drawback when it comes to background suppression since at HL-LHC one expects a lot of pile-up events where the produced muon has a charged meson in its vicinity. Another difficulty is that the neutrino produced in the meson decay is much softer compared to the $W$-boson decay. As can be seen in~\cref{fig:illustration}, neutrinos from $W$-bosons mainly produce displaced jet events with $E \!>\! 100~\gev$ in the rapidity range accessible at the HGCAL, while those produced in QCD events preferably distribute in the lower displaced jet energy range around $\sim 50~\gev$. However, in this energy range the background from soft neutral QCD is significantly larger. While in these heavy flavour processes the overall neutrino production rate is higher than in $W$ production, the enhanced background levels will most probably require a more complex search strategy including e.g.~jet shape variables. We leave a dedicated analysis of these processes for future work. 
 
\section{Conclusions and Outlook}\label{sec:results}

In this paper we have studied the potential of the future CMS HGCAL to detect scattering of muon-neutrinos produced in $W$-boson decays. 
In this proof-of-principle study we have demonstrated that with a suitable choice of simple analysis-level cuts, $\mathcal{O}(10)$ muon-neutrino scattering events can be collected during the HL-LHC. Simultaneously, the enormous expected background rates from neutral hadron scattering can be reduced to a similar level by this same set of cuts. This makes the study of neutrino scattering in the $100~\gev - 1~\tev$ energy range feasible at the HL-LHC, a kinematic range where very few data exists.

From the left panel of~\cref{fig:sigbkg} and the cut flow table shown in~\cref{tab:cutflow}, it can be seen that cutting on the energy carried away by the secondary muon produced in the neutrino scattering is vital for reducing the neutral hadron background. Demanding at least 33\% of the energy of the incoming neutrino to be carried away the muon decreases the amount of background events almost by a factor of $10^{4}$ while it only reduces the signal by $\sim17$\%.
Finally, the most important cut for background reduction is the jet energy correlated with the jet isolation cut. This is illustrated in the left panel of~\cref{fig:deltaR}. The right panel of~\cref{fig:sigbkg} summarises the final selection of signal and background events after all cuts applied, where the vertical dashed line illustrates the final jet energy cut on a  sufficiently \textit{isolated} high energy displaced jet. As can be seen in the last row of~\cref{tab:cutflow}, the jet energy cut reduces the number of background events by another factor of $10^{5}$. Thus, after all cuts applied  one expects to detect $\sim3.6$ muon-neutrino scattering events and $\sim3.5$ background events from neutral hadron scattering.

This corresponds to a signal-to-background ratio of $S/\sqrt{B}\sim 1.9$ and makes the first direct observation of neutrinos at a high-energy collider multipurpose experiment feasible if further improvements in the analysis are made. While we have already detected neutrinos with few $100~\gev$ energies in dedicated neutrino experiments at Tevatron, and anticipate to detect more at far foward neutrino experiments at the LHC, those neutrinos are produced in meson decays. The search strategy presented in this paper allows for the first direct observation of neutrinos produced in decays of $W$-bosons.

\bigskip

\textbf{Future improvements and applications -}  The search strategy presented in this work can be optimized by using e.g.~two-dimensional cuts. For example, in the left panel of~\cref{fig:deltaR} it can be seen that one in principle a better signal-to-background discrimination could be achieved by using a two-dimensional cut in the jet energy versus $R$ plane. Ultimately, this type of search could be even further improved by using  a fully-fledged multivariate analysis technique. In this context, one could also include more low-level features like shower shape variables to discriminate displaced hadronic from neutrino jets. For this purpose it will also be necessary to analyse the detector response in a full \texttt{Geant~4} simulation. 

Apart from the aforementioned qualitative improvements of the analysis, we have pointed out in~\cref{sec:nuscattering} that an orthogonal direction would be to search for neutrinos in decays of $B$ and $D$ mesons. A further interesting direction is to study the decay $W\to \nu_\tau \, (\tau \to \nu_\tau\, \mu\nu_\mu)$, where the prompt decay of the produced tau produces a muon and muon neutrino. If these are energetic enough they could pass the $W$ mass selection criterion and contribute to the muon neutrino signal. We leave this for further study.

Beyond the scope of SM neutrino interactions the type of search strategy proposed in this work can be applied to a vast class of (light neutral) new physics like neutrino NSIs, light mediators or DM. In this context,  it has recently been demonstrated that searches for appearing recoil jets have a unique sensitivity to hadronically coupled light pseudo-scalar DM~\cite{Bauer:2020nld}. Another scenario considers the production of dark sector states via muon interactions with the HCAL~\cite{Galon:2019owl}. Since these types of searches for appearing displaced jets target a complementary set of physics goals than standard LHC searches, its potential should be further studied in the future to optimally exploit the physics output of the LHC  program as a whole.

\vspace{-0.5cm}
\section*{Acknowledgements}
\vspace{-0.3cm}

The authors want to thank Martin Bauer and Tilman Plehn for early collaboration and valuable input on this project and Jonathan Feng, Max Fieg and Mauro Valli for feedback on the manuscript. Furthermore, the authors are very grateful to Jakob Salfeld-Nebgen for helpful discussions on the experimental details of this analysis. PF is funded by the UK Science and Technology Facilities Council (STFC) under grant ST/P001246/1. The work of FK is supported by the U.S.~Department of Energy under Grant No.~DE-AC02-76SF00515 and by the Deutsche Forschungsgemeinschaft under Germany’s Excellence Strategy - EXC 2121 Quantum Universe - 390833306. PR acknowledges financial support from the Fundação de Amparo à Pesquisa do Estado de São Paulo (FAPESP) under the contract 2020/10004-7.


\bibliography{literature}

\end{document}